\documentclass[format=acmsmall, review=false, screen=true, natbib=true]{acmart}

\usepackage[inline]{enumitem}
\usepackage{multirow}
\usepackage{tabularx}
\usepackage{xcolor}
\usepackage{colortbl}
\usepackage[normalem]{ulem}
\newcolumntype{C}[1]{>{\centering\arraybackslash}m{#1}}

\usepackage{caption}

\definecolor{darkgreen}{rgb}{0.01, 0.75, 0.24}

\setcopyright{rightsretained} 
\copyrightyear{2024}
\acmYear{2024}
\acmDOI{10.1145/3629981}
\acmJournal{TORS}       
\acmVolume{2}           
\acmNumber{2}           
\acmArticle{12}
\acmYear{2024}
\acmMonth{4} 

\begin{document}

\title{Generating Usage-related Questions for Preference Elicitation in Conversational Recommender Systems}

\titlenote{An early version of this work has been published as a late-breaking results paper at RecSys'21~\citep{Kostric:2021:RecSys}.}

\author{Ivica Kostric}
\affiliation{%
  \institution{University of Stavanger}
  \city{Stavanger}
  \country{Norway}
}
\email{ivica.kostric@uis.no}

\author{Krisztian Balog}
\affiliation{%
  \institution{University of Stavanger}
  \city{Stavanger}
  \country{Norway}
}
\email{krisztian.balog@uis.no}

\author{Filip Radlinski}
\affiliation{%
  \institution{Google}
  \country{UK}
}
\email{filiprad@google.com}

\begin{abstract}
  A key distinguishing feature of conversational recommender systems over traditional recommender systems is their ability to elicit user preferences using natural language.  Currently, the predominant approach to preference elicitation is to ask questions directly about items or item attributes.  
  Users searching for recommendations may not have deep knowledge of the available options in a given domain. As such, they might not be aware of key attributes or desirable values for them.
  However, in many settings, talking about the \emph{planned use} of items does not present any difficulties, even for those that are new to a domain.  In this paper, we propose a novel approach to preference elicitation by asking implicit questions based on item usage.   As one of the main contributions of this work, we develop a multi-stage data annotation protocol using crowdsourcing, to create a high-quality labeled training dataset. 
  Another main contribution is the development of four models for the question generation task: two template-based baseline models and two neural text-to-text models.  The template-based models use heuristically extracted common patterns found in the training data, while the neural models use the training data to learn to generate questions automatically.
  Using common metrics from machine translation for automatic evaluation, we show that our approaches are effective in generating elicitation questions, even with limited training data.  
  We further employ human evaluation for comparing the generated questions using both pointwise and pairwise evaluation designs. We find that the human evaluation results are consistent with the automatic ones, allowing us to draw conclusions about the quality of the generated questions with certainty.  Finally, we provide a detailed analysis of cases where the models show their limitations.
\end{abstract}



\copyrightyear{2022}

\maketitle

\section{Introduction}
\label{sec:intro}

Traditionally, recommender systems predict users' preference towards an item by performing offline analysis of past interaction data (e.g., click history, past visits, item ratings)~\citep{Gao:2021:AI}.  These systems often do not take into account that users might have made mistakes
in the past (e.g., regarding purchases)~\citep{Wang:2021:WSDM} or that their preferences change over time~\citep{Jagerman:2019:WSDM}.  Additionally, for some users, there is little historical data which makes modeling their preferences difficult~\citep{Lee:2019:KDD}.
A \emph{conversational recommender system} (CRS), on the other hand, is a multi-turn, interactive recommender system that can elicit user preferences in real-time using natural language~\citep{Jannach:2022:ACM}.  Given its interactive nature, it is capable of modeling dynamic user preferences and taking actions based on users current needs~\citep{Gao:2021:AI}.

One of the main tasks of a conversational recommender system is to elicit preferences from users. This is traditionally done by asking questions either about items directly or item attributes \citep{Christakopoulou:2016:KDD, Gao:2021:AI, Sepliarskaia:2018:RecSys, Christakopoulou:2018:KDD, Zhang:2018:CIKM, Chen:2012:User,Wu:2019:RecSys, Sun:2018:SIGIR, Lei:2020:KDD}.
Asking people to review individual recommendations to establish the characteristics of a single item they need, especially in a domain that they are not expert in, is particularly time consuming; therefore,
the research is commonly focused on the estimation and utilization of users preferences towards attributes~\citep{Gao:2021:AI}. Common to these approaches is that the user is explicitly asked about the desired values for a specific product attribute, much in the spirit of slot-filling dialogue systems~\citep{Gao:2018:SIGIR}.  For example, in the context of looking for a bicycle recommendation, we might have wheel dimensions or the number of gears as attributes in our item collection.  In this case, a system might want to ask a question like \emph{``How thick should the tires be?''} or \emph{``How many gears should the bike have?''}  However, ordinary users often do not possess this kind of attribute understanding, which might require extensive domain-specific knowledge.  Instead, they only know where or how they intend to use the item.  For example, a user might only be interested in using this bike for commuting but does not know what attributes might be good for that purpose.
The novel research objective of this work is to generate \emph{implicit} attribute questions for eliciting user preferences, related to the intended use of items.  This stands in contrast to explicit questions that ask about specific item attributes.

\begin{figure*}[t]
  \centering
  \captionsetup{width=.9\linewidth}
  \includegraphics[width=.9\linewidth]{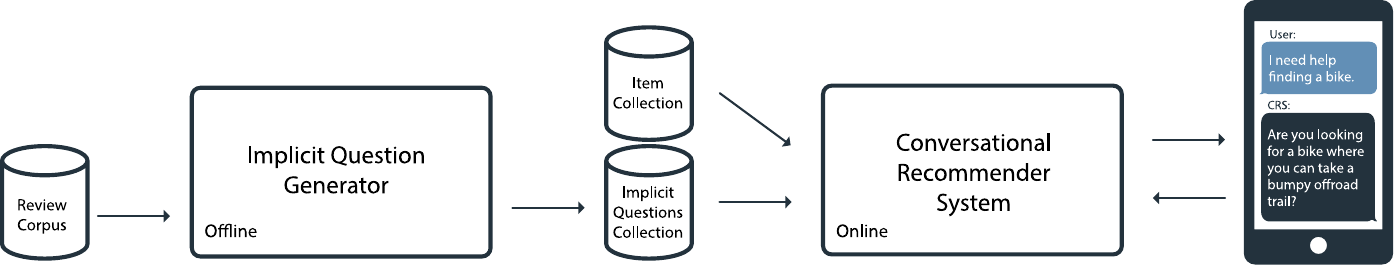}
  \caption{Conceptual system overview. Our focus in this paper is on the implicit question generator component.}
  \label{fig:crs_intro}
\end{figure*}

Our approach hinges on the observation that usage-related experiences are often captured in item reviews.  By identifying review sentences that discuss particular item features or aspects (e.g., \emph{``fat tires''}) that matter in the context of various activities or usage scenarios (e.g., \emph{``for conquering tough terrain''}), those sentences can then be turned into preference elicitation questions.
In our envisaged scenario, a large collection of implicit preference elicitation questions is generated offline, and then utilized later in real-time interactions by a CRS; see Fig.~\ref{fig:crs_intro} for an illustration.

In this paper, our focus is on the offline question generation part, whereas the actual item recommendation is left as a separate, downstream task to be addressed.
A main challenge associated with the question generation task is the collection of high-quality training data.
As our first contribution, we address the problem of creating a sentence-to-question dataset by developing a multi-stage data generation protocol.
It starts with \emph{candidate sentence selection}, which can be automated effectively based on part-of-speech tagging and simple linguistic patterns.
Then, we employ a multi-step manual data annotation process via crowdsourcing, which involves (1) question generation (given an input sentence, turn it into a question, if possible), (2) question validation (filtering the responses collected in the previous step), and (3) expanding question variety (producing paraphrased versions of the input questions). 
As our second contribution, we propose four \emph{question generation} models that, given a review as input, produce an implicit question in an end-to-end fashion. The simplest, template-based model uses the most common n-grams found in the training data to construct questions. The second model extends this template-based baseline by adding a classifier to discard non-applicable sentences before generating a question. The last two are neural models we fine-tuned for this particular task, from a pre-trained checkpoint of a sequence-to-sequence model for text generation~\citep{Raffel:2020:JMLR}. The difference between the latter two lies in what is taken as the input---the first model uses heuristically extracted sentences, while the second one uses an entire review.
The evaluation of our proposed approach is done against held-back test data using standard metrics for \emph{automatic evaluation} of text generation (BLEU, ROUGE, and METEOR).  Additionally, we evaluate the task in terms of Accuracy, i.e., whether a question can be constructed based on the given input.
In \emph{human evaluation} we measure the effectiveness and the capability of our model to generate questions that are suitable for preference elicitation, can be answered easily, and are grammatically correct.  The evaluation is performed both in pointwise and pairwise fashion, using a 5-point Likert scale. 
We find that all evaluations results (both automatic and  two flavors of human evaluation) point to the same conclusions: that our proposed neural models outperform the strong template-based baseline. There are advantages to both neural models: the sentence-based model generates questions of slightly higher quality, while the review-based one has the advantage of being an end-to-end model with a simpler architecture.

In summary, our main contributions in this paper are as follows:
\begin{itemize}
    \item Introduce the novel task of eliciting preferences in CRSs via usage-related questions.
    \item Develop a multi-stage data annotation protocol using crowdsourcing for collecting high-quality ground truth data.
    \item Introduce two template-based and two neural approaches for generating usage-related questions based on a corpus of item reviews.
    \item Develop human evaluation protocols, conduct both automatic and manual evaluation of the proposed approaches, and perform an extensive analysis of results.
\end{itemize}
The resources developed in this paper (crowdsourced dataset and question generation model) are made publicly available at \url{https://github.com/iai-group/tors2023-crs-questions}.

\section{Related Work}
\label{sec:related}

The focus of this work is preference elicitation via natural language in the context of conversational recommender systems. In this section, we discuss related work on conversational recommender systems, preference elicitation, and question generation. 

\subsection{Conversational Recommender Systems}

Static recommendation models predict users' preferences based on their previous interactions with the system.  Some of the more common early approaches include collaborative filtering (CF)~\citep{Sarwar:2001:WWW}, logistic regression (LR)~\citep{Nelder:1972:RSS} and gradient boosting decision tree (GBDT)~\citep{Blanco:2013:ISWC}.
The availability of datasets on user behavior data (e.g., click history, visit logs, ratings on items\footnote{https://grouplens.org/datasets/movielens/}$^,$\footnote{https://www.baltrunas.info/context-aware}) has inspired, in recent years, the development of more sophisticated neural models such as neural factorization machines (NFM)~\citep{He:2017:SIGIR} or graph convolutional networks (GCN)~\citep{Ying:2018:KDD}.
A significant drawback of static recommenders is that they treat recommendation as a \emph{one-shot} interaction process under the assumption that the user's preferences lie in historical data.  However, this does not hold in cases where there are no past observations~\citep{Lee:2019:KDD}.  This is often the case in scenarios where the user has not interacted with the system (cold-start problem) or in the case with high-involvement products (i.e., products that customers do not buy frequently and tend to invest more time and effort when selecting them)~\citep{Jannach:2022:ACM}.  \citet{Wang:2021:WSDM} note that data on clicks and purchases could be misleading, because a large portion of clicks do not lead to purchases, and when they do, users might have regretted their choice.  Furthermore, the user's preferences might change over time \citep{Jagerman:2019:WSDM} and capturing their past interactions can lead to recommendations that are no longer relevant.
To deal with short-term but dynamic preferences, \emph{session-based recommenders} have emerged and received considerable attention in recent years~\citep{Wang:2021:ACM}.  These algorithms provide recommendations solely based on the user's interactions during a continuous period of time (i.e., a session).

A \emph{conversational recommender system} (CRS)
helps users reach their recommendation-oriented goals via multi-turn conversation~\citep{Jannach:2022:ACM}. While they share the goal of recommending items to users with traditional, static recommender systems, they do so by eliciting the detailed and current user preferences interactively in real-time. 
In contrast to session-based recommenders, where user preferences are implicit and inferred from interactions, users explicitly express their preferences here using natural language.  
Additionally, a CRS can provide explanations for the suggested items and process user feedback on the recommendation.
While there are many open issues around CRSs, \citet{Gao:2021:AI} identified the following five as primary challenges:
\begin{itemize}
    \item \emph{Question-based User Preference Elicitation.} The challenge is to generate questions that elicit as much information as possible and to use the provided information to make better recommendations. Two main lines of research are item-based~\citep{Zhao:2013:CIKM, Christakopoulou:2016:KDD, Sepliarskaia:2018:RecSys} and attribute-based preference elicitation~\citep{Zhang:2018:CIKM, Lei:2020:WSDM}. Both approaches try to answer the questions of what to ask and how to adjust the recommendation based on user response.
    \item \emph{Multi-turn Conversational Recommendation Strategies.} The main challenge is to balance continued question asking to reduce preference uncertainty and provide recommendations using the least number of conversation turns.
    \item \emph{Natural Language Understanding and Generation.} One of the hardest challenges in CRSs is to communicate like a human~\citep{Gao:2021:AI}.  Commonly, this involves providing a recommendation list directly or incorporating recommended items in a rule-based natural language template~\citep{Gao:2018:SIGIR, Habib:2020:CIKM, Zhang:2020:KDD}.  Recently, end-to-end frameworks have been proposed to understand users' intents and generate readable, fluent, and meaningful natural language responses~\citep{Li:2018:NIPS}.
    \item \emph{Trade-offs between Exploration and Exploitation.} The dynamic nature of CRSs allows them to actively explore unseen items to capture user preferences. However, users generally have limited time and energy to interact with the system, therefore systems need to balance exploration with exploitation to make accurate recommendation.
    \item \emph{Evaluation and User Simulation.} The complexity of evaluating CRSs comes from the emphasis on user experience during interactions. Systems need to be evaluated both on the turn and on the conversation level.  While static recommenders can utilize large quantities of historical data to evaluate models, obtaining large number of user interactions to evaluate CRS is expensive~\citep{Huang:2020:RecSys}.  Therefore, user simulation-based evaluation has been identified as a promising direction~\citep{Zhang:2020:KDD, Afzali:2023:WSDM}.
\end{itemize}

\noindent
In this paper, we focus on question-based user preference elicitation and natural language generation. That is, we provide novel answers to questions \emph{what to ask} and \emph{how to ask}.

\subsection{Preference Elicitation}

Commonly, preference elicitation questions target either items or their attributes.
Typical of early studies on CRSs, \emph{item-based elicitation} approaches to ask for users' opinions on an item itself, using a combination of methods from traditional recommender systems, such as collaborative filtering, with user interaction in real time~\citep{Zhao:2013:CIKM, Wang:2019:IEEE}. These systems continuously recommend items and refine the recommendations based on user feedback.
In case of \emph{choice-based methods}, users are presented with two or more items. In every turn, the recommendation is updated based on the selected choice. The selection of items may be approached as an optimization problem using a static preference questionnaire method~\citep{Sepliarskaia:2018:RecSys}.
Another line of research is using probabilistic, multi-armed bandit algorithms that maximize the cumulative expected reward over some fixed number of rounds. There is an inherent exploration-exploitation trade-off in these systems where exploration refers to acquiring information about arms, while exploitation is optimizing for the immediate reward in the current round~\citep{Christakopoulou:2016:KDD, Wang:2019:IEEE}. This method has a natural setup in the CRS setting where items can be seen as arms and rounds as the conversation turns.

Asking about items directly can be inefficient, as large item sets would require several conversational turns and in turn increase the likelihood of users losing interest~\citep{Gao:2021:AI}.
Alternatively, \emph{attribute-based elicitation} aims to predict the next attribute to ask about.  It is often cast as a sequence-to-sequence prediction problem, lending itself naturally to sequential neural networks~\citep{Hochreiter:1997:Neural, Cho:2014:EMNLP}.
There has been an effort to create large datasets consisting of human conversations that can be used as training data. However, non-conversational data is often leveraged, especially when there is a lack of relevant information in the recorded dialogues~\citep{Jannach:2022:ACM}.
\citet{Christakopoulou:2018:KDD} propose a question \& recommendation (Q\&R) method, utilize data from a non-conversational recommendation system, and develop surrogate tasks to answer questions: \emph{What to ask?} and \emph{How to respond?} To answer the first question, they develop a surrogate task where the goal is to predict the next likely topic a user would be interested in, based on recently watched videos. 
The second question is answered by predicting what video the user would be most interested in, based on the most relevant predicted topic.
A similar approach of training a sequential neural network on non-conversational data is taken by \citet{Zhang:2018:CIKM}, who convert Amazon reviews into artificial conversations.  Sentences with aspect-value pairs are extracted from reviews and serve as utterances in one round of conversation. The extracted aspect-value pairs are modeled as user information needs.
The assumption is that the earlier aspect-value pairs appear in the review, the more important they are to the user, and thus should be prioritized as questions.  Additionally, they develop a heuristic trigger to decide whether the model should ask about another attribute or recommend an item.
The drawback of these systems is they have no way of modeling the rejection of recommendations by the user, since the goal is to fit historical data as it happened. Furthermore, it is not possible to determine the reason behind the user interaction, i.e., why the user chose that particular item~\citep{Gao:2021:AI}.

Another way to elicit preferences is in the form of \emph{critiques}, i.e., feedback on attribute values of recommended items~\citep{Chen:2012:User}.
For example, if the recommendation is for a \emph{phone}, a critique might be \emph{``not so big''} or \emph{``something cheaper.''}
Such methods often employ heuristics as elicitation tactics~\citep{Luo:2020:SIGIR, Luo:2020:WWW}.
In recent work, \citet{Balog:2021:SIGIR} study the problem of robustly interpreting unconstrained natural language feedback on attributes.
Our work differs from prior efforts in that we do not ask about specific attribute values directly, but instead ask indirect questions related to the planned use of an item.

To help interactively search and navigate the space of item, \emph{facet-based selection} is a commonly used interaction paradigm, especially in e-commerce~\citep{Tunkelang:2009:Book}.
Facets correspond to a particular way of grouping items, based on attribute-value combinations.  For a given item category, facets may be identified by domain experts or sorted dynamically in order to allow for a quick
drill-down of the results~\citep{Vandic:2017:IEEE}.
Our work may be seen as a different way of clustering items, around item usage.  However, different from facet selection, there is no linear constraint on a single facet---item usage maps to a subset of the attribute space, without the user necessarily knowing what the facets are.  In practice, item selection often involves balancing a trade-off, e.g., a bike that is practical for daily usage and can be taken off-road occasionally.  This type of selection  can be done based on usage, but not with facets/attributes.

\subsection{Question Generation}

While there is research on end-to-end frameworks to enable CRSs to both understand user intentions as well as generate fluent and meaningful natural language responses~\citep{Li:2018:NIPS}, the predominant approach is still to use templates or construct the utterances using predefined language patterns~\citep{Gao:2021:AI}. 
In recent years, the broader field of dialogue systems has brought forth two additional strands of research applicable to CRSs as well: retrieval-based and generation-based methods~\citep{Manzoor:2021:RecSys}.
Instead of relying on a handful of templates, \emph{retrieval-based methods} utilize a large collection of possible responses.  The basic approach to retrieving the appropriate response is based on some notion of similarity between the user query and candidate responses, with the simplest being inner product~\citep{Wu:2019:SIGIR}.
\emph{Generation-based methods} in dialogue systems are typically based on sequence-to-sequence modeling.  These models are usually trained on a hand-labeled corpus of task-oriented dialogue~\citep{Budzianowski:2018:EMNLP}.
Due to the limited amount of training data, \emph{delexicalization} is used to increase the generality of the systems.  It is the process of disassociating specific words from the lexicon by replacing them in the training set with generic placeholders.
The sequence-to-sequence model is then trained to produce a delexicalized sentence (utterance skeleton) as output. To get the final sentence, the output utterance is \emph{relexicalized} based on user need~\citep{Jurafsky:2020:Book}.
Although retrieval-based approaches have been explored to a lesser extent than generation-based methods, their potential to leverage large, existing dialogue datasets to provide contextually relevant and high-quality responses has been demonstrated, resulting in an improved conversational user experience~\citep{Manzoor:2021:RecSys, Manzoor:2022:IS}.
Our proposed approach shares elements of both of retrieval-based and generation-based methods: it generates questions using a sequence-to-sequence model and stores them in a collection that can be queried using retrieval-based methods. However, the task we focus on is fundamentally different. Namely, we are concerned with preference elicitation through the generation of implicit questions based on item usage, rather than simply responding to user queries or generating dialogue. This renders existing approaches inadequate for our task. 

The problem of preference elicitation is also related to that of clarification of information needs in information-seeking scenarios. 
When searching for information, user queries are often ambiguous, faceted, or incomplete.  To improve the user satisfaction, systems may decide not to provide an answer (e.g., based on their estimated confidence in the results) but instead proactively ask the user questions to clarify their needs~\citep{Aliannejadi:2019:SIGIR}.  This is especially important in conversational information seeking scenarios, where the system can return only limited number of results due to the limited bandwith user interface.  Similar to research in CRS, existing approaches to generating clarifying questions include retrieval-based methods~\citep{Aliannejadi:2019:SIGIR, Rao:2018:ACL, Yang:2020:WWW} and generation-based methods~\citep{Wang:2018:ACL, Rao:2019:NAACL}.  Our work differs from this line of work in that instead of clarifying an already expressed need, we are trying to elicit a new user information need.

\subsection{Sequence-to-Sequence Modeling}

The task of sequence-to-sequence models is to generate a sequence of output tokens conditioned on the input sequence.  To generate high-quality output, transfer learning has proved to be a powerful technique.  In transfer learning, a model is first pre-trained on a data-rich task, then fine-tuned on a downstream task.  Early implementations used recurrent neural networks~\citep{peters:2018:NAACL}, however, in recent years, the Transformer architecture is more commonly used~\citep{Vaswani:2017:NIPS}.
Within the Transformer framework, three main variants emerged: encoder-only, decoder-only, and encoder-decoder models. Encoder-only models, like BERT~\citep{Devlin:2019:NAACL}, are mainly used for classification.
On the other hand, for text generation tasks, decoder-only~\citep{Radford:2019:} and encoder-decoder models~\citep{Raffel:2020:JMLR, Lewis:2020:ACL} are often employed.
One of the main differences of the two variants used for generation, apart from the architecture, is in the pre-training regime. Encoder-decoder models are generally pre-trained using causal masked token prediction, where a number of tokens in any position of the input sequence are masked and the model predicts the masked tokens based on the context. Decoder models, on the other hand, are pre-trained using a next token prediction strategy based on the input sequence plus the tokens predicted thus far.
Both training regimes are conducted in an unsupervised fashion, and the goal is to learn language syntax and semantics, and store that information in the model weights.
In this work, we apply sequence-to-sequence models to the question generation task with the goal of generating  usage-related questions using different inputs (sentences or entire reviews) as context.
\section{Approach}
\label{sec:approach}

Our objective is to understand users' needs with minimal cognitive effort on their part.  To overcome the shortcomings associated with item-based elicitation (large item space and slow narrowing of the recommendation candidates) and attribute-based elicitation (domain-specific knowledge required), we propose asking usage-related questions instead.  These should be easier for users to answer and can thus lead to a better conversational user experience. 

As a first step toward that objective, in this work, we focus on the generation of implicit elicitation questions---implicit in the sense that we ask users about the intended use of items as opposed to soliciting the values of specific attributes. To generate usage-related questions, we leverage review corpora under the assumption that reviewers bring attention to item usage, where or how an item was used, and whether or not it was suitable for the intended purpose.
 We want to identify item uses that occur sufficiently frequently and could be converted to a good question to present to a new user.
Item review datasets tend to be very large, with both the number of items and reviews in the thousands or even millions, making manual labeling the entire dataset extremely expensive~\citep{Liao:2021:arXiv}.
To overcome this, we develop automated approaches that can take a review as input and generate one or multiple preference elicitation questions out of that, if it is possible.
We present multiple methods with an increasing degree of automation:
\begin{itemize}
  \item We start with a \emph{template-based} baseline approach that follows a pipeline of steps: first splitting reviews to sentences, then selecting sentences that mention some item-related activity or usage, and finally turning those sentences to questions using a pre-defined pattern (Section~\ref{sec:approach:template}).  This approach will always yield a question if the input sentence mentions an activity.
  \item Our second baseline extends the template-based approach by adding a classifier that is tasked with selecting only those sentences that could be converted to good questions. Otherwise, it still uses templates to construct questions from the selected sentences (Section~\ref{sec:approach:template-classification}).
  \item Next, we introduce a \emph{neural sentence-based} approach, which still operates on the sentence level, but handles activity detection and question generation in an end-to-end manner using a large pre-trained language model (Section~\ref{sec:approach:neural-sentence}).
  \item Finally, we present a \emph{neural review-based} method, which takes an entire review as input and generates a review questions from that, if it is possible (Section~\ref{sec:approach:neural-review}).
\end{itemize}
Figures~\ref{fig:TQG}--\ref{fig:NRQG} present schematic overviews of the different methods. 

\subsection{Baseline 1: Template-based Question Generation}
\label{sec:approach:template}

\begin{figure*}[t]
  \centering
  \includegraphics[width=1\linewidth]{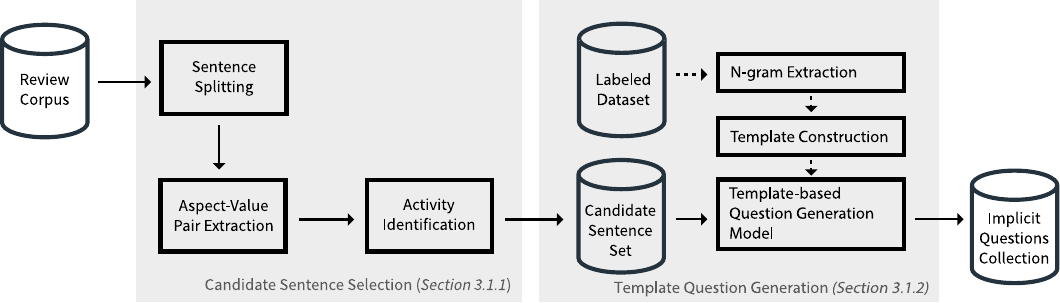}
  \caption{Components of our template-based question generation system.}
  \label{fig:TQG}
\end{figure*}

Figure~\ref{fig:TQG} illustrates the components of our \emph{template-based question generation} (TQG) approach.

\subsubsection{Candidate Sentence Selection}
\label{sec:approach:sentence}

We identify sentences that describe some item feature or aspect and mention some activity or usage.  For example:

\begin{center}
  \vspace*{0.5em}
  \small
  \texttt{The \(\overbrace{\text{fat}}^{value}\) \(\overbrace{\text{tires}}^{aspect}\) are perfect for \(\overbrace{\text{conquering tough terrain}}^{usage/activity}\).}
  \vspace*{0.5em}
\end{center}

\paragraph{Aspect-Value Pair Extraction}
\label{sec:approach:sentence:aspect}

An aspect in this context is a term that characterizes a particular feature of an item \citep{Lu:2011:WWW} (e.g., \emph{wheel}, \emph{seat} or \emph{gear} are aspects of a bicycle).  Value words are terms that describe an aspect (e.g., a \emph{wheel} might be \emph{large} or \emph{small}, a \emph{seat} can be \emph{hard} or \emph{comfortable}). Here, we extract all sentences that mention some aspect-value pair for a given category of items, using phrase-level sentiment analysis proposed by~\citet{Zhang:2014:SIGIRa, Zhang:2014:SIGIR}.
The motivation for this step stems from the assumption that an activity or usage can be mapped to a particular aspect of an item.\footnote{This concerns future utilization of responses given to these elicitation questions, where the CRS might want to map activities to specific attribute values.}

\paragraph{Activity Identification}
\label{sec:approach:sentence:activity}

In this step, the goal is to classify sentences that mention some item-related activity or usage. Inspired by \citet{Benetka:2019:CHIIR}, our approach revolves around using \emph{part-of-speech} (POS) analysis and rules of the English language.  We filter for the preposition \emph{for} followed by a verb in progressive tense heuristically, by looking for \emph{-ing} endings (e.g., \emph{for commuting}, \emph{for hiking}).  
This choice is driven by our intuition and was verified by manually inspecting a sample of the data.
Note that there might be other formulations that describe activity or usage. Our goal is not to extract all possible sentences containing mentions of activity or usage; a high recall approach would likely come at the cost of a larger fraction of false positives. Instead, we focus on achieving high precision.

\subsubsection{Question Generation}
\label{sec:approach:question}

The main motivation for this step is generating natural-sounding questions that are easy for users to understand and answer, without needing any additional context.
Consider the sentence \emph{``The fat tires are perfect for conquering tough terrain.''}  An example of converting it to a yes or no usage-related question might be \emph{``Would you like a bike that is perfect for conquering tough terrain?''}
We approach this task using a template-based method, which is a common approach in CRS question generation~\citep{Gao:2018:SIGIR, Habib:2020:CIKM, Zhang:2020:KDD}.

There are many possible ways of articulating questions.  To ensure that they are as natural-sounding as possible, we develop our template based on actual questions that humans formulated from review sentences.  That is, we assume the presence of a training dataset consisting of sentence-question pairs, and inspect the most commonly appearing n-grams from the questions in that dataset.  Specifically, in our training dataset (cf. Section~\ref{sec:data}), we observe the following as the most frequent question pattern:
\begin{center}
  \vspace*{0.5em}
  \small
  \texttt{Are you looking for a [category] that is great for [usage]?}
  \vspace*{0.5em}
\end{center}
An example question, based on this template, would be: \emph{``Are you looking for a bike that is great for commuting?''}

Note that not all candidate sentences that pass our selection heuristic are viable for conversion to a question, e.g., \emph{``Thank you so much for coming up with such a great product.''} This sentence is too vague and does not mention any action or usage for the item, and thus should be labeled as not applicable (N/A).  However, the simple template-based approach is not capable of making such distinction and would generate a question regardless.

\subsection{Baseline 2: Template-based Question Generation with Classification}
\label{sec:approach:template-classification}

With our second model (TQG+CLS), we address some of the limitations of the first baseline model. Specifically, we aim to avoid generating questions that would not help with recommendations, because they would either be trivially answered affirmatively or would not make it easier to make a recommendation. Before generating a question, we classify the sentence as applicable or not applicable. If the sentence is not applicable, we do not generate a question.
To achieve this, we train a transformer-based classifier to predict whether a sentence is applicable or not. We choose RoBERTa~\citep{Liu:2019:arXiv}, a high-performant BERT-based transformer model.
The input to the model is:
\begin{center}
  \vspace*{0.5em}
  \small
    \texttt{input\_seq = <cls> [sentence] <eos>} 
  \vspace*{0.5em}
\end{center}
where \texttt{<cls>} and \texttt{<eos>} are special tokens that mark the beginning and end of the sequence, respectively. \texttt{cls} is used as an input to a simple linear and a softmax layer, whose output gives us probability distribution over the two classes: applicable and not applicable.

\subsection{Neural Sentence-based Question Generation}
\label{sec:approach:neural-sentence}

\begin{figure*}[t]
  \centering
  \includegraphics[width=1\linewidth]{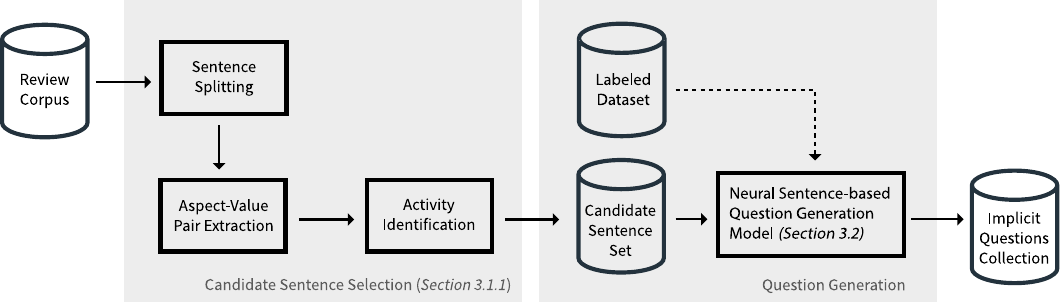}
  \caption{Components of our neural sentence-based question generation system. The approach is similar to that of the template-based question generation, but instead of creating rigid templates, the model learns question patterns from the entire dataset automatically using a neural model.}
  \label{fig:NSQG}
\end{figure*}

In our third model, \emph{neural sentence-based question generation} (NSQG), depicted in Fig.~\ref{fig:NSQG}, we further address some of the limitations of the template-based approach. 
First, similarly to Baseline 2, we want to produce relevant questions for recommendations by avoiding those that are either easily answered affirmatively or do not contribute to facilitating a recommendation.  For example, instead of generating the question \emph{``Do you want a grill that is good for grilling certain things?,''} we want the model to output a special \texttt{[N/A]} (not applicable) token.
Second, we would want to generate a richer variety of natural-sounding questions.

Learning to generate questions is done by fine-tuning a large, pre-trained, sequence-to-sequence language model.
There are two main benefits of using transfer learning from a pre-trained model. First, it increases the learning speed; as both syntax and semantics of the English language are already learned, there are fewer things the model needs to learn.  %
Second, it reduces the amount of labeled data needed to train models to high performance.  %
Specifically, we use T5~\citep{Raffel:2020:JMLR} as it has shown competitive performance across a variety of language generation tasks (e.g., conversational query rewriting~\citep{Lin:2020:arXiv} and document re-ranking~\citep{Pradeep:2021:arXiv}), and can be used for both N/A-classification and generation, where N/A-classification is posed as a text-to-text problem.
We form the input to the T5 model as follows:
\begin{center}
  \vspace*{0.5em}
  \small
    \texttt{input\_seq = Ask question: [category] <sep> [sentence] <eos>} 
  \vspace*{0.5em}
\end{center}
 where ``\texttt{Ask question:}'' is a task-specific prefix, and \texttt{<sep>} and \texttt{<eos>} are the separation and end-of-sequence tokens, respectively. Considering that T5 was pre-trained on various tasks, a task-specific prefix is used to specify which task the model should perform. The output of the model is either a question or the \texttt{[N/A]} token.
 
We employ state-of-the-art techniques when performing model inference.  Specifically, we use temperature-controlled stochastic sampling with top-$k=25$ and top-$p=0.90$ (nucleus) filtering.  Top-\emph{k} sampling restricts the sampling to consider only the $25$ most likely next tokens.  However, since some distributions from which tokens are sampled are flat while others are sharp, fixed \emph{k} sampling is not optimal.  To mitigate this shortcoming, nucleus filtering restricts the number of considered tokens to the minimum number of tokens whose total probability sums to $p$.

\subsection{Neural Review-based Question Generation}
\label{sec:approach:neural-review}

The main motivation behind our last model, \emph{neural review-based question generation} (NRQG), is to simplify the process of question generation. 
The model is an extension of the previous sentence-based question generation (NSQG) model, except the input being an entire review instead of a single sentence.

\begin{figure*}[t]
  \centering
  \includegraphics[width=.5\linewidth]{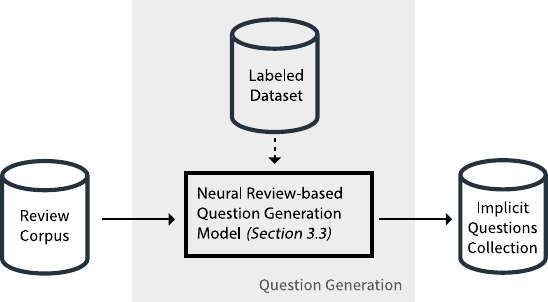}
  \caption{Components of our neural review-based question generation system. The model drastically simplifies inference as we do not rely on heuristics to extract candidate sentences, but take entire reviews as input to generate questions.}
  \label{fig:NRQG}
\end{figure*}

On a high level, review-based question generation is a two-step process: a text extraction step, to identify a review sentence, followed by a text generation step where the sentence is ``translated'' into a question.
That is, meaningful usage-related information first needs to be found in a longer text and then be converted into a question.
While sentence-based approaches use a heuristic for the first step and either a template (TQG) or a trained neural model (NSQG) for the second, with NRQG we simplify the pipeline considerably.
Using a neural architecture allows us to perform both steps jointly by fine-tuning a large pre-trained language model in an end-to-end fashion, as illustrated in Fig.~\ref{fig:NRQG}.  

The input to the NRQG model follows a similar structure to NSQG, except we replace the sentence with a review.
\begin{center}
  \vspace*{0.5em}
  \small
    \texttt{input\_seq = Ask question: [category] <sep> [review] <eos>} 
  \vspace*{0.5em}
\end{center}
While many neural models have an input limitation, usually of 512 tokens, T5 has no such limitation. However, long input sequences drastically slow down generation, and the common practice is to avoid them. In our experiments, only a handful of reviews were longer than 512 tokens with the longest having 2000 tokens. If long reviews were common, a solution to processing longer reviews would be to use the same approach by splitting the reviews into manageable-sized chunks.
Another consideration is dealing with reviews that mention multiple possible uses for an item.  This is not a common scenario, and there are indeed no examples of such cases in our dataset.  Therefore, we make the simplifying assumption that at most a single question may be generated from one review.

It is important to note that while we use an existing model for text generation in both sentence-based and review-based models, obtaining high-quality labeled data for fine-tuning the model is a challenge on its own.  As one of the contributions of this paper, we develop a multi-step data collection protocol using crowdsourcing, which we discuss in Section~\ref{sec:data:crowdsourcing}.  Furthermore, while NRQG simplifies the modeling part considerably, it still relies on high-quality training data.  The candidate sentence selections described in Section~\ref{sec:data:reviews} is thus instrumental to facilitating data collection.  In our experiments (in Section~\ref{sec:results}), we will analyze the impact of the pre-trained language model (i.e., number of parameters) as well as the amount of training data available for fine-tuning.

\section{Data Collection}
\label{sec:data}

\begin{figure*}[t]
  \centering
  \includegraphics[width=1\linewidth]{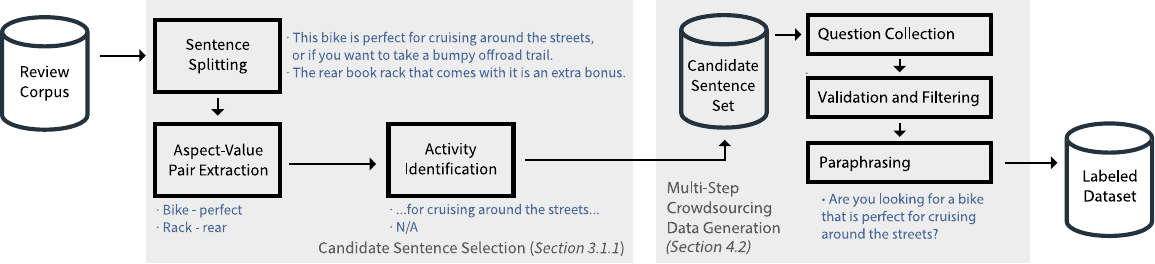}
  \caption{Data collection pipeline, consisting of automatic candidate sentence extraction based on linguistic patterns and multi-step manual data annotation via crowdsourcing.}
  \label{fig:data_collection}
\end{figure*}

This section describes the process of creating our dataset, which consists of a set of review sentences and either (i) a corresponding set of five preference elicitation questions or (ii) the label not applicable (N/A) for each.  The data collection pipeline is shown in Figure~\ref{fig:data_collection}.

\subsection{Candidate Sentence Selection}
\label{sec:data:reviews}

The starting point for getting the candidate sentences are the Amazon review and metadata datasets~\citep{Ni:2019:EMNLP},\footnote{https://nijianmo.github.io/amazon/index.html} where item reviews from Amazon are extracted along with product metadata information such as \emph{title}, \emph{description}, \emph{price}, and \emph{categories}. There are, in total, \(233.1\)M reviews about \(15.5\)M products.  Due to the sheer dataset size, we focus our research on three main categories: \emph{Home and Kitchen}; \emph{Patio, Lawn and Garden}; and \emph{Sports and Outdoors}. From these (40M reviews), we further sub-select 12 diverse subcategories (referred to as \emph{categories} henceforth): \emph{Backpacking Packs}, \emph{Tents}, \emph{Bikes}, \emph{Jackets}, \emph{Vacuums}, \emph{Blenders}, \emph{Espresso Machines}, \emph{Grills}, \emph{Walk-Behind Lawn Mowers}, \emph{Birdhouses}, \emph{Feeders}, and \emph{Snow Shovels}. This narrowed down the number of reviews to 989k.

Sentence splitting and \emph{aspect-value} pair extraction is performed using the Sentires toolkit~\citep{Zhang:2014:SIGIRa, Zhang:2014:SIGIR}.\footnote{https://github.com/evison/Sentires}  This step discards many non-viable sentences.
The remaining ones are POS-tagged using the Stanford NLP toolkit~\citep{Manning:2014:ACL}.
Finally, we filter for sentences that match our activity detection heuristic (\emph{``for + [verb in progressive tense]''}). After this step, we were left with 14,140 reviews.
Our sentence selection process is designed to favor precision over recall, and was validated by manual inspection of the results.
Upon completion of the crowdsourcing tasks (described in Section~\ref{sec:data:crowdsourcing}), we find that over 75\% of the selected sentences can be turned into questions.
This shows that our simple method can indeed identify candidate sentences with relatively high precision.

Our final \emph{candidate sentence set} contains approximately 100 sentences per category. An exception is the \emph{Birdhouses} category, where only 15 candidate sentences are found due to the size of that category. In total, the candidate set consists of 1,098 sentences over 12 categories.

\subsection{Question Generation using Crowdsourcing}
\label{sec:data:crowdsourcing}

Crowdsourcing was done on the Amazon Mechanical Turk (AMT) platform in three steps. The task was available to workers with \(95\%\) approval rate and with at least 1,000 approved human intelligence tasks (HITs).

\subsubsection{Step 1: Question Collection}

Crowd workers are given a review sentence (describing some aspect or use for a product) and a product category as input, and tasked with rewriting it into a question or marking it as not applicable.
They are specifically instructed to formulate a question that a salesperson or a recommender agent might ask a customer, such that it is a standalone question that can simply be answered with yes/no.
For every input sentence, we collected responses from three different workers. Sentences found non-applicable by at least two workers are set as N/A. The task was re-run if a single worker responded with N/A. This process resulted in approximately 2,600 sentence-question pairs.

\subsubsection{Step 2: Validation and Filtering}

Next, we validate all responses (i.e., generated questions) for applicable sentences collected in Step 1 using crowdsourcing.
We employ three different workers in Step 2, who are requested to answer four multiple-choice questions:
(1) \emph{Is the question grammatically correct?} [Yes/No]
(2) \emph{Can the question be answered by yes or no?} [Yes/No]
(3) \emph{Does the question mention any trait or use for a product?} [Yes/No]
(4) \emph{Who is most likely to ask this question in a sales setting?} [Buyer/Salesperson/Neither].
Generated questions that are found invalid by all three workers on a single aspect or at least two workers on at least two aspects are automatically rejected.  Those that are marked invalid on multiple aspects but do not fall into the former category are manually checked by an expert annotator (one of the authors).  All other questions are approved. Steps 1 and 2 were run multiple times until all questions were resolved.

\subsubsection{Step 3: Expanding Question Variety}

Our main motivation for expanding the question variety is to add new ways of asking implicit questions.  To this end, we task a new set of workers to paraphrase the questions we obtained and validated in Steps 1 and 2.  Each worker receives all three versions of the questions from Step~~1 as input and is asked to produce a new (paraphrased) question that expresses the same meaning.  Note that this set of workers do not get to see the original sentences, only the questions generated from them by other workers.  For each set of three questions, two additional paraphrases were collected.  Considering that generating paraphrases proved to be a much simpler task than generating questions from review sentences, no additional quality assurance steps were necessary.

\subsection{Final Dataset}
\label{sec:data:final_dataset}

Out of the 1,115 candidate sentences, 277 were labeled as non-applicable (not containing relevant usage-related information), which is below \(25\%\).  This shows that our high-precision approach to selecting candidate sentences is effective.  We note that our sentence selection method works better for some categories than for others.  The fraction of viable sentences ranges between \(52\%\) (\emph{Espresso machine} category) to \(84\%\) (\emph{Backpacking pack} category).
For the remaining 838 sentences, a total of five questions are generated, three based on the candidate sentence and two via paraphrasing.  Table~\ref{tab:example_amt} shows two example sentences from our dataset.

\begin{table}[t]
    \centering
    \caption{Example sentence-question pairs from our dataset.}
    \label{tab:example_amt}
    \footnotesize
    \begin{tabularx}{1\textwidth}{|l|X|}
        \hline
        \textbf{Category} & Blender \\
        \hline
        \textbf{Sentence} & Great for making smoothies with frozen fruit. \\
        \hline
        \textbf{Generated questions}
        	& -~~ Are you looking for a blender that's great for making smoothies with frozen fruit? \\
        	& -~~ Would you be interested in a blender that is great for making smoothies with frozen fruit? \\
        	& -~~ Are you interested in a blender for making smoothies with frozen fruit? \\
        
        \hline
        \textbf{Paraphrases}
        	& -~~ Do you want a blender that's great for making smoothies with frozen fruit? \\
        	& -~~ Would you like a blender that is great for making smoothies with frozen fruit? \\
           \hline
           \multicolumn{2}{c}{~} \\
            \hline
            \textbf{Category} & Snow shovel \\
        \hline
        \textbf{Sentence} & This product is excellent for doing the job \\
        \hline
        \textbf{Generated questions}
        	& n/a \\
            & n/a \\
            & n/a \\
        \hline
        \textbf{Paraphrases}
        	&  \\
        	&  \\
        \hline
    \end{tabularx}
\end{table}

\section{Experimental Setup}
\label{sec:evaluation}

This section presents the experimental setup for the three methods explored in this paper. We evaluate our models using standard automatic metrics for evaluating text generation. We also perform human evaluation via a set of crowdsourcing studies to assess question quality across multiple dimensions.

\subsection{Question Generation}
For our neural approaches (NSQG and NRQG), we train \emph{small}, \emph{base}, and \emph{large} T5 models, which vary in the number of layers, self-attention heads, and the dimension of the final feedforward layer. The difference is shown in the number of parameters in Table~\ref{tab:results:size}.  We use \(80\%\) of the data for training, while the rest is test data.  In our training, we employ teacher forcing~\citep{Williams:1989:Neural}, regularization by early stopping~\citep{Morgan:1989:NIPS}, and adaptive gradient method AdamW~\citep{Loshchilov:2022:ICLR} with linear learning rate decay.
For each sentence, we have either N/A or a set of reference questions as ground truth.

\subsection{Automatic Evaluation}

We evaluate question generation as a classification task in terms of Accuracy (detecting N/A), and as a machine translation task, where the set of human-generated questions serve as reference translations.
Specifically, we report on BLEU-4, which uses modified n-gram precision up to 4-grams~\citep{Papineni:2002:ACL}, and ROUGE-L, a recall-based metric based on the longest common subsequence~\citep{Lin:2004:ACL}. Additionally, we report METEOR, which has been found to have a better correlation with human judgments compared to BLEU and ROUGE~\citep{Lavie:2007:StatMT}. It does this by considering word stems, WordNet synonyms, and paraphrases in addition to n-gram overlap.

While evaluating sentence-based models (TQG and NSQG) is straightforward, there is a detail we have to consider when generating and evaluating questions using the review-based (NRQG) method.  Each review may contain multiple sentences mentioning usage that could potentially be used to generate questions.  However, in our dataset we do not have any such instances (i.e., no two sentences happen to come from the same review).  While this is not by design, intuitively it makes sense that people do not discuss multiple usages of an item within a single review.
Therefore, we can evaluate the NRQG model exactly the same way we evaluate the TQG and NSQG models---that is, for each input review we expect a single generated question.  The generated question is then compared to the ground truth questions in the dataset.  

\subsection{Human Evaluation}

Recent studies have shown that automatic measures often have a low correlation with human judgement~\citep{Sai:2022:ACM, Mairesse:2010:ACL, Liu:2016:EMNLP, Callison-Burch:2006:EACL, Sekulic:2021:ICTIR}.  To thoroughly investigate the differences between our question generation models, we additionally evaluate them by human assessors.  Human evaluation of natural language generation is most commonly done with respect to a single dimension, however, it has been observed that there are many aspects of language generation that cannot be captured in a single metric~\citep{derLee:2021:CSL}.  In our work, we consider three quality dimensions: grammar and fluency, usability, and answerability.

We compare the generated questions both on their own (\emph{pointwise} evaluation) and relative to each other (\emph{pairwise} evaluation) with the help of crowd workers.
Research suggests that pairwise comparison might be more reliable \citep{Carterette:2008:ECIR}, however, the cost of evaluation increases with the number of models. 
Another reason for performing pointwise evaluation is that it yields an \emph{absolute} measure by averaging over a set of questions.  Pairwise evaluation, on the other hand, can only establish a \emph{relative} ordering between two approaches.
Nevertheless, both absolute and relative measurements can be insightful and we are particularly interested in seeing if the observations we can draw from them are in alignment. In both cases, we focus on three different aspects of the questions, which in combination describe their \emph{naturalness}.

An important aspect of conversational systems is to generate fluent, coherent, and grammatically correct utterances~\citep{Anand:2020:Dagstuhl}. Therefore, the first evaluation dimension focuses on \emph{grammar and fluency.} In pointwise evaluation, we ask \emph{``Is the question fluent and grammatically correct?,''} while in the pairwise case we ask \emph{``Which question is more fluent and grammatically correct?''} Another important aspect when generating questions that are supposed to elicit user preferences is  \emph{usefulness}. \citet{Rosset:2020:WWW} introduce the concept of usefulness in conversational search and describe it as a measurement for how well a suggestion leads the user to useful information. In our case, we are trying to evaluate how useful the question is in making a good recommendation. In other words, does answering the question help with giving a better recommendation?
We ask (pointwise) \emph{``If you were making a recommendation for a friend, would knowing the answer be useful for you to make a better recommendation?''} or (pairwise) \emph{``If you were making a recommendation for a friend, the answer to which question would be more useful for you to make a better recommendation?''} The final aspect we explore is that of \emph{answerability}, i.e., how easy or difficult it is for the user to answer the question. Is the question ambiguous or straightforward? For example, a question \emph{``Are you looking for a snow shovel that is extremely good snow shovel for Wyoming?''} might be easy to answer for most people living in Wyoming or if we know what the typical winter is like there.  However, this kind of question is very specific and difficult to answer for most people outside Wyoming.  In pointwise evaluation, we ask \emph{``Would you expect someone looking for a recommendation to be able to answer this question easily?,''} while in pairwise evaluation, we ask \emph{``Which question is easier to answer when looking for a recommendation?''}

\begin{figure*}[t]
  \centering
  \includegraphics[width=.8\linewidth]{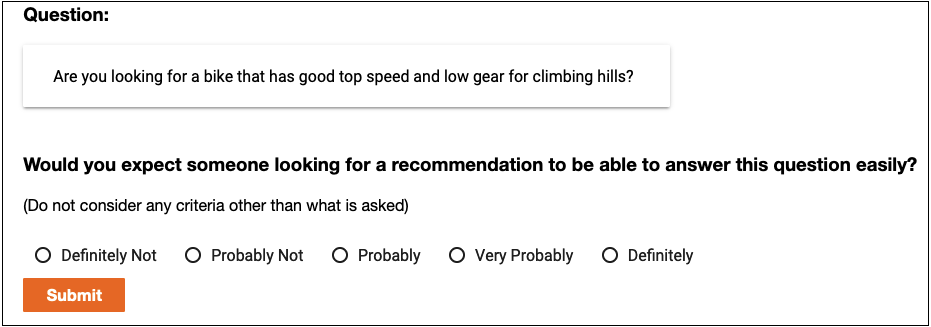}
  \caption{Example question for pointwise evaluation. The specific task addresses the answerability of the question in the context of providing better recommendations.}
  \label{fig:pointwise}
\end{figure*}

\begin{figure*}[t]
  \centering
  \includegraphics[width=.9\linewidth]{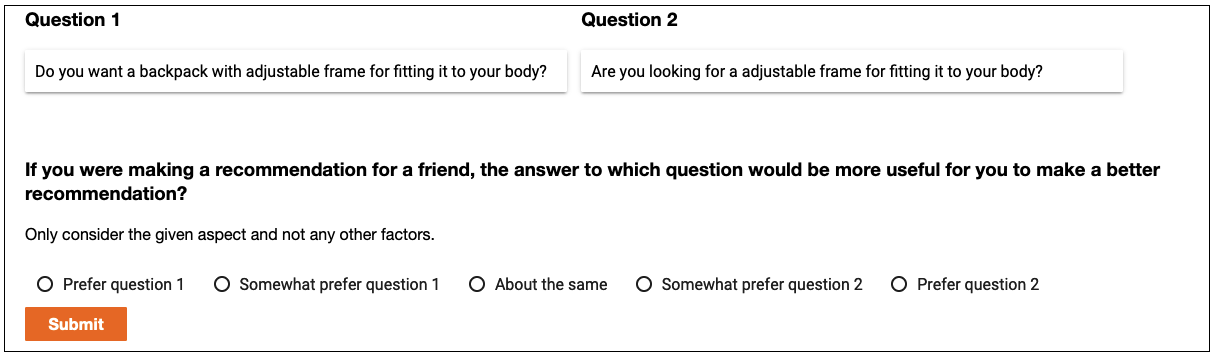}
  \caption{Example question for pairwise evaluation. The specific task addresses the usefulness of the presented question.}
  \label{fig:pairwise}
\end{figure*}

In pointwise evaluation we solicit answers on a 5-point Likert scale. An example can be seen in Fig.~\ref{fig:pointwise} where the responses range from ``definitely not'' to ``definitely.''
In pairwise evaluation we also employ a 5-point scale, where the two ends of the spectrum correspond to strong preferences for each question, with gradually weaker preferences in between and ``no preference'' in the middle. An example pairwise evaluation task is shown in Fig.~\ref{fig:pairwise}.

We use crowdsourcing on Amazon Mechanical Turk to evaluate the generated questions in our study.  Each question is annotated by three different workers, all based in the United States or Great Britain, with a minimum approval rate of 95\%, and a minimum number of accepted HITs 1000.  We take the mean of the three annotations as the final score for each question. 

\section{Results and Analysis}
\label{sec:results}

The main research question we wish to answer with our experiments is the following: \emph{Given a review from a corpus, can item-usage questions for preference elicitation be automatically generated?} To address this question, we break the problem down into more specific research questions:
\begin{itemize}
    \item \textbf{RQ1} Can neural models generate more natural questions when compared with template-based baselines?
    \item \textbf{RQ2} How does (a) the size of the pre-trained language model and (b) the volume of available training data affect the performance of the neural models?
\end{itemize}
Specifically, given a review as input, our approaches should either generate a question or label it as not applicable (N/A) if a usage-related question cannot be generated or where the generated question would not be useful.

\subsection{Automatic Evaluation}
\label{sec:results:results}

\begin{table}[t]
    \centering
    \caption{Performance comparison of different question generation models: template-based (TQG), neural sentence-based (NSQG), and neural review-based (NRQG).  All models utilize all available training data (i.e., five questions or N/A per sentence).  The best scores for each measure are in boldface. The symbols $^*$ and $^+$ denote statistically significant improvements over the two baselines, respectively (p-value $< 0.05$). Statistical significance for accuracy is measured using McNemar's test, while for BLEU, ROUGE, and METEOR we use paired bootstrap resampling~\citep{Koehn:2004:EMNLP}.}
    \label{tab:results}
    \begin{tabularx}{\linewidth}{l@{\extracolsep{\stretch{1}}}llll}
        \toprule
        \textbf{Model} & \textbf{N/A Accuracy} & \textbf{BLEU-4} & \textbf{ROUGE-L} & \textbf{METEOR} \\
        \midrule
        Baseline 1 (TQG)      & 0.728 & 0.604 & 0.723 & 0.418    \\
        Baseline 2 (TQG+CLS)       & \textbf{0.870$^*$} & 0.607 & 0.727 & 0.420    \\
        NSQG      & 0.858$^*$ & \textbf{0.730$^{*+}$} & \textbf{0.806$^{*+}$} & \textbf{0.494$^{*+}$} \\
        NRQG      & 0.832$^*$ & 0.684$^{*+}$ & 0.769$^{*+}$ & 0.466$^{*+}$     \\
        \bottomrule
    \end{tabularx}
\end{table}

First, we compare the neural models with the two baseline models using automatic evaluation (RQ1). We train \emph{T5-large} for both NSQG and NRQG as it was found to be the most effective model for the task. The results are reported in Table~\ref{tab:results}.  We find that both neural models significantly outperform the template-based models on all generation evaluation metrics. This is expected, as neural models are capable of using both syntax and semantics present in the original sentence when generating questions. Unsurprisingly, they significantly outperform Baseline 1 on the classification task as well. Baseline 2 achieves the highest accuracy, suggesting that a dedicated classifier performs better than a general-purpose model on the classification task.
We also observe that the sentence-based (NSQG) model outperforms the review-based one (NRQG) on all metrics.  This is unsurprising as the NSQG model has a simpler task to perform, as it receives the already extracted candidate sentences as input.  Note that the accuracy of the review-based model is much higher than that of the template-based model, and only slightly worse than that of the sentence-based model, which suggests that despite the larger (and arguably noisier) input, the model can predict with high accuracy if useful questions can be generated.

\subsection{Human evaluation}

\begin{table}[t]
    \centering
    \caption{Pointwise evaluation of our three models for each quality dimension, on a scale of 1 to 5. The best scores for each measure are in boldface. The symbols $^*$ and $^+$ denote statistically significance improvements over the two baselines, respectively (p-value $< 0.05$), measured using a non-parametric Mann-Whitney U test.}
    \label{tab:results:pointwise}
    \begin{tabularx}{\linewidth}{l@{\extracolsep{\stretch{1}}}lll}
        \toprule
        \textbf{Model} & \textbf{Grammar and Fluency} & \textbf{Usefulness} & \textbf{Answerability} \\
        \midrule
        Baseline 1 (TQG)         & 3.69     & 3.48     & 3.71   \\
        Baseline 2 (TQG+CLS)       & 3.85$^*$     & 3.67$^*$     & 3.86   \\
        NSQG        & \textbf{4.02$^*$}     & \textbf{3.81$^{*+}$}     & \textbf{3.98$^*$}   \\
        NRQG        & 3.80     & 3.73$^*$    &  3.93$^*$  \\
        \bottomrule
    \end{tabularx}
\end{table}

\begin{figure}[t]
    \includegraphics[width=\linewidth]{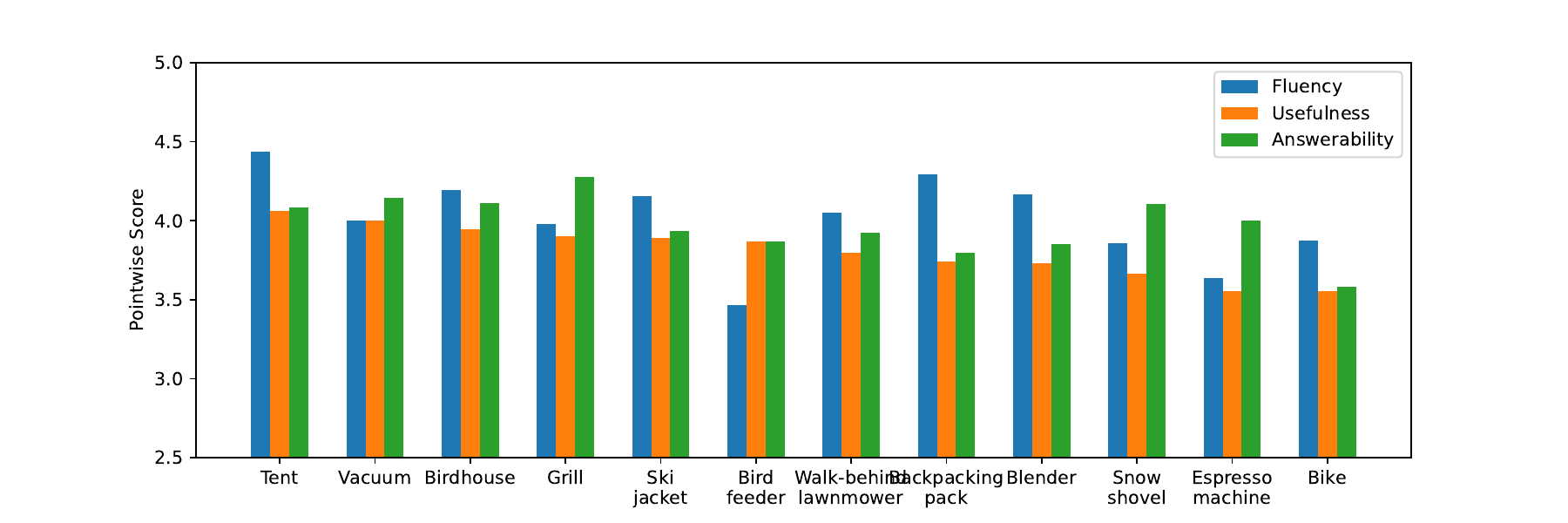}
    \caption{Pointwise evaluation of the NSQG model per category. Catrgories are sorted by the usefulness score.}\label{fig:results:pointwise_categories}
\end{figure}

\begin{table}[t]
    \centering
    \small
    \caption{Pairwise evaluation according to different quality dimensions. The main values are percentages of how often the model in the row wins over the model in the column. The value in brackets are percentages of how often the model is strongly preferred.}
    \label{tab:results:pairwise}
    \begin{tabularx}{.31\linewidth}{C{.67cm}|C{.74cm}C{.74cm}C{.74cm}}
        \multicolumn{4}{c}{\emph{Grammar and fluency}} \\
        \hline
         & \multicolumn{3}{c}{\textbf{Wins over}} \\
         & \textbf{TQG} & \textbf{NSQG} & \textbf{NRQG} \\
        \hline
	\hfil
	\textbf{TQG}	&	--	&	{\cellcolor[rgb]{0.607,0.847,0.727}} 31\% {\scriptsize (10\%)}	&	{\cellcolor[rgb]{0.474,0.795,0.771}} 38\% {\scriptsize (16\%)}\\
	\textbf{NSQG}	&	{\cellcolor[rgb]{0.358,0.731,0.81}} 44\% {\scriptsize (20\%)}	&	--	&	{\cellcolor[rgb]{0.513,0.812,0.758}} 36\% {\scriptsize (15\%)}\\
	\textbf{NRQG}	&	{\cellcolor[rgb]{0.513,0.812,0.758}} 36\% {\scriptsize (17\%)}	&	{\cellcolor[rgb]{0.662,0.868,0.711}} 28\% {\scriptsize (10\%)}	&	-- \\
    \end{tabularx}
    \quad
    \begin{tabularx}{.31\linewidth}{C{.67cm}|C{.74cm}C{.74cm}C{.74cm}}
        \multicolumn{4}{c}{\emph{Usefulness}} \\
        \hline
         & \multicolumn{3}{c}{\textbf{Wins over}} \\
         & \textbf{TQG} & \textbf{NSQG} & \textbf{NRQG} \\
        \hline
	\textbf{TQG}	&	--	&	{\cellcolor[rgb]{0.75,0.902,0.75}} 22\% {\scriptsize (6\%)}	&	{\cellcolor[rgb]{0.706,0.885,0.731}} 25\% {\scriptsize (7\%)}\\
	\textbf{NSQG}	&	{\cellcolor[rgb]{0.513,0.812,0.758}} 36\% {\scriptsize (17\%)}	&	--	&	{\cellcolor[rgb]{0.706,0.885,0.731}} 25\% {\scriptsize (7\%)}\\
	\textbf{NRQG}	&	{\cellcolor[rgb]{0.513,0.812,0.758}} 36\% {\scriptsize (16\%)}	&	{\cellcolor[rgb]{0.737,0.897,0.744}} 23\% {\scriptsize (7\%)}	&	-- \\
    \end{tabularx}
    \quad
    \begin{tabularx}{.31\linewidth}{C{.67cm}|C{.74cm}C{.74cm}C{.74cm}}
        \multicolumn{4}{c}{\emph{Answerability}} \\
        \hline
         & \multicolumn{3}{c}{\textbf{Wins over}} \\
         & \textbf{TQG} & \textbf{NSQG} & \textbf{NRQG} \\
        \hline
	\textbf{TQG}	&	--	&	{\cellcolor[rgb]{0.585,0.839,0.734}} 32\% {\scriptsize (11\%)}	&	{\cellcolor[rgb]{0.513,0.812,0.758}} 36\% {\scriptsize (16\%)}\\
	\textbf{NSQG}	&	{\cellcolor[rgb]{0.419,0.765,0.79}} 41\% {\scriptsize (18\%)}	&	--	&	{\cellcolor[rgb]{0.546,0.824,0.747}} 34\% {\scriptsize (16\%)}\\
	\textbf{NRQG}	&	{\cellcolor[rgb]{0.452,0.783,0.779}} 39\% {\scriptsize (19\%)}	&	{\cellcolor[rgb]{0.568,0.832,0.74}} 33\% {\scriptsize (14\%)}	&	-- \\
    \end{tabularx}
\end{table}

The questions generated by the four models are also evaluated using human assessors along three dimensions: grammar and fluency, usefulness, and answerability.  
The pointwise evaluation results are presented in Table~\ref{tab:results:pointwise}.  Overall, all models score above average ($>$ 3) along all evaluation dimensions.  The neural models outperform Baseline 1 when comparing \emph{grammar and fluency}; the differences are significant for NSQG.  This is expected as the characteristic property of using large pre-trained language models is their capability to use grammar correctly.  When constructing templates using the most frequent n-grams, we have no guarantees of fluency or adherence to grammatical rules.  However, it is interesting to note that grammar is adequate (i.e., scoring 3 or greater) in over 80\% of the test cases and that Baseline 2 significantly improves it.  In both \emph{usefulness} and \emph{answerability}, the neural models perform similarly, with the review-based (NRQG) being only slightly worse than the sentence-based (NSQG) model.  They both significantly outperform Baseline 1, which likely follows from the fact that these models can accurately determine when not to generate a question and predict N/A instead.

Figure~\ref{fig:results:pointwise_categories} shows the breakdown of the pointwise evaluation across all 12 product categories for the sentence-based neural model (NSQG).  We see the scores are above average (i.e., above 3) for all categories on all three dimensions.  
Of the three dimensions, the scores for \emph{grammar and fluency} are the highest overall, as well as for most categories.  Interestingly, there is still a large variance between different categories, with the categories \emph{Bird feeder} and \emph{Espresso machine} having the lowest scores, and \emph{Tent} and \emph{Backpacking pack} highest scores.
The categories \emph{Bike}, \emph{Espresso machine}, and \emph{Snow shovel} have the lowest scores in terms of \emph{usefulness}.  It suggests that the model should label sentences as N/A more often for those categories. 

The pairwise evaluation follows the same patterns as the pointwise evaluation.  In all cases, annotators prefer the outputs of the NSQG model, followed by the NRQG model.  The biggest distinction between the template-based and neural models is seen in \emph{usefulness}, where the annotators prefer the neural models, often strongly so, in the vast majority of cases.  There is almost no distinction between the neural models. However, NSQG is a clear favorite in the other two dimensions (i.e., \emph{grammar and fluency} and \emph{answerability}).

To answer our main research question, we conclude that overall, we can generate high-quality questions according to both automatic and human evaluation.  Furthermore, based on human evaluation experiments, the neural models generate more natural questions compared to the template-based baselines (RQ1).

\subsection{Model Size}

\begin{table}[t]
    \centering
    \caption{Performance of the sentence-based question generation (NSQG) model using different pre-trained language models that are fine-tuned on all available training data (i.e., five questions or N/A per sentence).  The best scores for each measure are in boldface.}
    \label{tab:results:size}
    \begin{tabularx}{\linewidth}{l@{\extracolsep{\stretch{1}}}rrrrr}
        \toprule
        \textbf{Model} & \textbf{\#Parameters} & \textbf{N/A Accuracy} & \textbf{BLEU-4} & \textbf{ROUGE-L} & \textbf{METEOR} \\
        \midrule
        T5-small       & 60.5 M                & 0.724     & 0.716            & \textbf{0.810}             &   \textbf{0.497}   \\
        T5-base        & 222 M                 &   0.819          & 0.693 & 0.794 & 0.493     \\
        T5-large       & 737 M                 & \textbf{0.858} & \textbf{0.730} & 0.806 & 0.494    \\
        \bottomrule
    \end{tabularx}
\end{table}

Next, we explore what effect the size of the pre-trained language model has on the performance of neural question generation (RQ2a).  Specifically, we fine-tune three T5 models of different sizes when employing neural sentence-based question generation (NSQG). Table~\ref{tab:results:size} shows the results in terms of non-applicability classification (Accuracy) and question generation (BLEU, ROUGE, and METEOR).
The model size does not have a large impact on the question generation task. The difference, however, is more pronounced for non-applicability (N/A) detection than for question generation.  Detecting N/A is one of the most important parts of the pipeline since question generation quality heavily depends on only converting useful item-usage sentences to questions.  Furthermore,  while there is a trade-off between model size and accuracy, note that the planned usage is to generate questions offline and store them as a question collection. Thus, efficiency is not the main concern in this scenario. For this reason, we conclude that larger pre-trained models yield more effective questions.

\subsection{Training Data Volume}

\begin{figure}[t]
    \begin{tabular}{ll}
        \includegraphics[width=.48\linewidth]{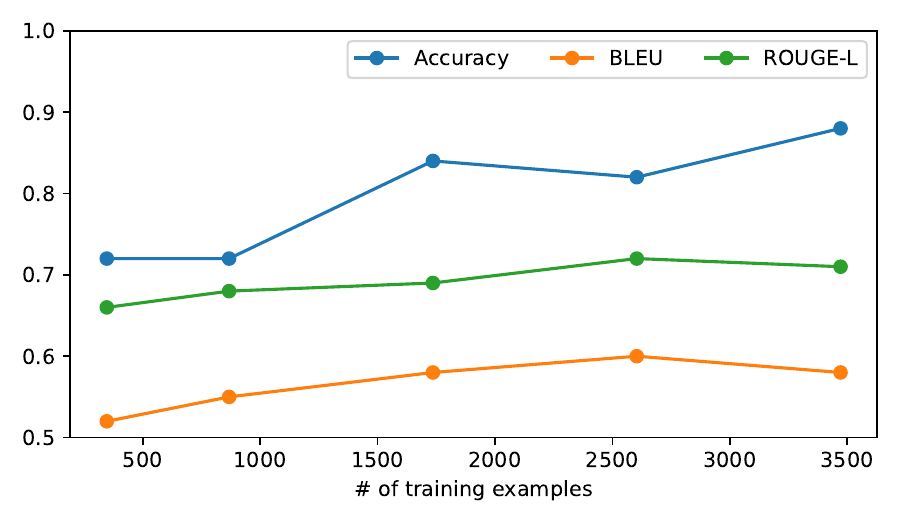} &
        \includegraphics[width=.48\linewidth]{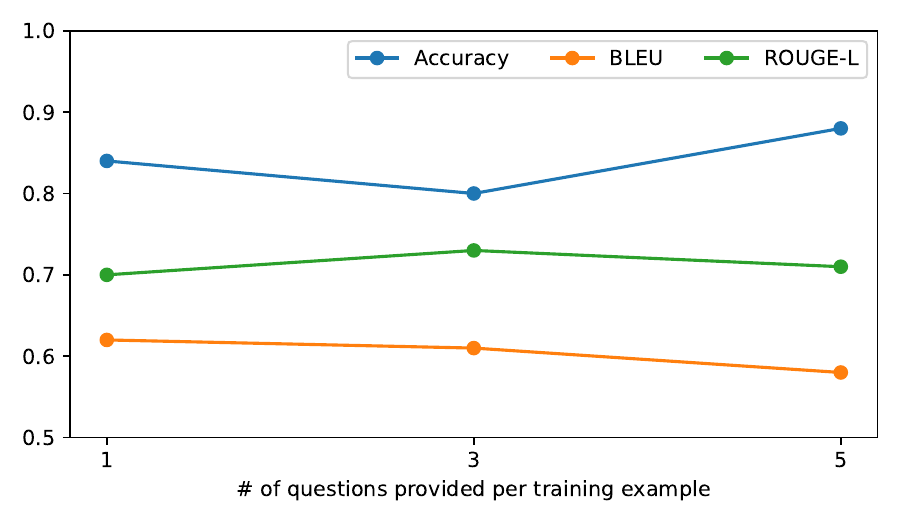}     \\
    \end{tabular}
    \caption{Model performance (T5-large) with sentence-based (Left) or question-based (Right) training data reduction for the T5-large version of the NSQG model.
    } \label{fig:results}
\end{figure}

We further investigate how the amount of training data affects model performance (RQ2b) by considering different ways and degrees of data reduction.  As before, we use the best-performing NSQG model for this experiment, i.e., T5 large.
In \emph{sentence-based} data reduction, shown in Fig.~\ref{fig:results} (Left), only a subset of the available sentences is used for training (using all available questions corresponding to those sentences).
We observe a drop in accuracy when we reduce the amount of training data to 25\% or lower (i.e., less than 1000 training samples), while question generation performance is less severely affected.
In \emph{question-based} data reduction, shown in Fig.~\ref{fig:results} (Right), we split the dataset based on the number of questions available for each sentence.  We consider using a single question (1), the three initially generated questions (3), and the three initial questions plus the two paraphrases (5).  We find that reducing the number of questions has surprisingly little effect.
This suggests that it is more beneficial to collect a small number of questions for a larger set of sentences than vice versa.

\subsection{Success/Failure Analysis}
\label{sec:results:analysis}

\begin{table}[t]
    \centering
    \caption{Examples of question generation outputs for all four models}.
    \label{tab:examples}
    \footnotesize
    \begin{tabularx}{1\textwidth}{|l|X|}
    
        \hline
        \textbf{Pattern} & \emph{Generic questions} \\
        \hline
        \multirow{3}{*}{\textbf{Ground truth}}
        
        & -~~ n/a \\
        & -~~ n/a \\
        & -~~ n/a \\
        \hline
        \multirow{3}{*}{\shortstack[l]{\textbf{TQG}\\\textbf{TQG+CLS}}}
        & -~~ Are you looking for a grill that is great for grilling certain things--not good for everything but what is? \\ 
        & -~~ Are you looking for a vacuum that is great for something to keep my floors clean -- not 'eat off of' clean? \\
        & -~~ Are you looking for a espresso machine that is great for making espresso drinks? \\
        \hline
        \multirow{3}{*}{\textbf{NSQG}}
        & -~~ Do you need a grill that is good for grilling certain things? \\
        & -~~ Are you looking for a vacuum to clean your floors? \\
        & -~~ Do you want an espresso machine that is good for making espresso drinks? \\
        \hline
        \multirow{3}{*}{\textbf{NRQG}}
        & -~~ Are you looking for a grill that is perfect for satay and quick grilling using smaller amounts of charcoal? \\
        & -~~ Are you looking for a vacuum cleaner that can keep your floors clean? \\
        & -~~ Are you looking for an espresso machine that is good for making espresso drinks? \\
        \hline
        \multicolumn{2}{c}{~} \\
        
        \hline
        \textbf{Pattern} & \emph{Complex questions} \\
        \hline
        \multirow{3}{*}{\textbf{Ground truth}}
        & -~~ n/a \\
        & -~~ Do you need a backpack with a good size for traveling on an airplane or going on a camping trip for a few days or packing for a few days trip? \\
        \hline
        \multirow{4}{*}{\shortstack[l]{\textbf{TQG} \\ \textbf{TQG+CLS} }}
        & -~~ Are you looking for a espresso machine that is great for something higher quality that'll last longer and has more features or if you'd like an introductory machine that makes delicious espresso that's much more affordable? \\
        & -~~ Are you looking for a backpacking pack that is great for traveling on an airplane (come up with ways to shrink it by removing the frame and using the straps) or going on a camping trip for a few days or packing for a few days trip? \\
        \hline
        \multirow{3}{*}{\textbf{NSQG}}
        & -~~ Are you looking for an espresso machine that has a lot of features or if you'd rather spend a lot of money up front for something higher quality that'll last longer and has more features? \\
        & -~~ Are you looking for a good size backpack that can be easily shrunk by removing frame and be used with straps? \\
        \hline
        \multirow{3}{*}{\textbf{NRQG}}
        & -~~ Are you looking for an espresso machine that lasts longer and has more features? \\
        & -~~ Are you looking for a backpack that is a good size for traveling on an airplane or going on a camping trip for a few days or packing for a few days trip? \\
        \hline

        \multicolumn{2}{c}{~} \\
        \hline
        \textbf{Pattern} & \emph{Success} \\
        \hline
        \multirow{3}{*}{\textbf{Ground truth}}
        & -~~ Would you like a spacious backpacking pack? \\
        & -~~ Are you looking for a grill that you can take on camping trips,even the long ones? \\
        & -~~ Do you want a perfect tent for backpacking? \\
        \hline
        \multirow{3}{*}{\shortstack[l]{\textbf{TQG} \\ \textbf{TQG+CLS}}}
        & -~~ Are you looking for a backpacking pack that is great for everything i would need for a three day isolation and more? \\
        & -~~ Are you looking for a grill that is great for taking on camping trips (even long ones)?\\
        & -~~ Are you looking for a tent that is great for backpacking? \\
        \hline
        \multirow{3}{*}{\textbf{NSQG}}
        & -~~ Are you in need of a backpack that has more than enough room for everything? \\
        & -~~ Are you looking for a grill that is perfect for long distance camping trips? \\
        & -~~ Are you looking for a tent that is perfect for backpacking? \\
        \hline
        \multirow{3}{*}{\textbf{NRQG}} 
        & -~~ Are you in need of a backpack that is in great shape and has more than enough room for everything? \\
        & -~~ Are you looking for a grill that is perfect for camping trips? \\
        & -~~ Are you looking for a tent that is perfect for backpacking? \\
        \hline
        
    \end{tabularx}
\end{table}

A closer look at specific sentence-question pairs reveals two patterns that leave room for future improvement; Table~\ref{tab:examples} shows examples of these using the template-based baseline models (TQG and TQG+CLS) and the best-performing variants of the neural models (NSQG and NRQG).
We find that some of the generated questions are \emph{too generic} (Table~\ref{tab:examples}, top block).  These are correct in terms of grammar and structure, but unsuitable for eliciting meaningful user preferences, e.g., \emph{``Do you need a grill that is good for grilling certain things?''}
Instead of returning N/A (which is indeed the corresponding response in our dataset), the model generated a question that is so vague and generic that it is hard to think of a scenario where it would not be answered affirmatively.  Interestingly, the review-based model in this scenario utilized another part of the input instead of the heuristically extracted sentence, which sentence-based models operate on, to generate a more useful question \emph{``Are you looking for a grill that is perfect for satay and quick grilling using smaller amounts of charcoal?''}
The second pattern concerns \emph{complex questions} (Table~\ref{tab:examples}, middle block) that ask about more than one usage or activity, e.g., \emph{``Are you looking for a backpacking pack that is a good size for traveling on an airplane or going on a camping trip for a few days or packing for a few days trip?''} This question is too complex and unlikely to elicit any meaningful information without the user having to elaborate which options they agree with and which they do not.  Such questions should instead be split into several simpler ones where it is both easier to interpret the question and to answer it.  Note that crowd workers were not instructed to simplify complex questions, therefore it is not surprising that is what the model has learned.
We also include examples of successes (Table~\ref{tab:examples}, bottom block) where all three models generate valid questions. We notice that for shorter inputs, all three models generate useful and grammatically correct questions that are easy to answer.  Since the template-based model is directly dependent on the structure of the input sentence, in some cases it does not produce a fluent question, e.g., \emph{``Are you looking for a backpacking pack that is great for everything i would need for a three day isolation and more?''} However, the meaning is still understandable even if the usefulness is limited of such an over-specified question.

\section{Conclusion and Future Directions}
\label{sec:concl}

In this paper, we have studied the question of how a conversational recommender system can solicit user's needs through natural language by using indirect questions about how the wanted product will be used.  This contrasts with most prior work that considers how to directly ask about desired product attributes.  
We have developed, evaluated, and compared four models used on the task: two template-based and two neural models.  In each case, the start is a corpus of reviews, and the goal is to generate preference elicitation questions, if possible.  We show that all four models effectively extract relevant information from reviews (with high precision), and transform it into useful questions.  For the sentence-based models, sentences containing usage-related statements are identified heuristically, while the review-based model works end-to-end.  The generated questions from all models are of high quality, with the neural models achieving higher scores in the automated evaluation and also being preferred by human annotators.

\emph{\textbf{Utilization.}}
We emphasize that this work focuses on this first stage of recommendation in a conversational setting, eliciting the user's needs in a natural and engaging way. The most important future direction is determining how answers to these questions should best be leveraged for the task of generating recommendations, once the user's need is understood. Here, we anticipate that sentence embedding techniques would likely to be effective.  Second, as this work builds on top of large language models, language safety is a key consideration warranting further study before our approach could be used in practice.  Nevertheless, during experimentation, we did not observe concerning language nor hallucinations.  We also note that the offline question generation process lends itself well to even manual control over the language model output. 

\emph{\textbf{Limitations.}}
We focus on generating high-quality questions (precision) as opposed the having an extensive coverage of the possible item uses (recall).
We do not address the aspect of question diversity explicitly. Instead, it is assumed that human-created reviews naturally cover the different ways a given item is used. 
Determining whether the coverage of usage-related questions is sufficient for a given category would be an interesting direction for further investigation.

\emph{\textbf{Generalizability.}}
Our approach may be employed in other domains where items are associated with a certain activity. For instance, in a movie recommendation scenario, a statement like ``This movie was perfect for watching with my kids,'' could provide valuable usage-related insights, as could similar statements in the domain of travel, food, or restaurants.
Our approach may be less applicable in contexts where the items do not lend themselves to specific activities or usage scenarios, e.g., news recommendation.

\bibliographystyle{ACM-Reference-Format}
\bibliography{references}

\end{document}